\documentclass[aps,twocolumn]{revtex4}
 \pdfoutput=1
\usepackage{graphicx}
\usepackage{bm}
\usepackage{braket}
\usepackage{pifont}
\usepackage{amsmath}
\usepackage{amsfonts}
\usepackage{amssymb}
\usepackage{graphicx}
\usepackage[FIGBOTCAP]{subfigure}
\usepackage{color}
\newcommand {\mrm}[1] {\mathrm{#1}}
\begin{document}

\title{Continuous Variable Quantum Repeaters} 

\author{Josephine Dias}
\email{josephine.dias@uqconnect.edu.au}
\author{Tim Ralph}

\affiliation{Centre for Quantum Computation and Communication Technology$,$ School of Mathematics and Physics$,$ University of Queensland$,$ St. Lucia$,$ Queensland 4072$,$ Australia}

\date{\today}

\begin{abstract}
We propose a quantum repeater for continuous variable (CV) quantum optical states. Our repeater relies on an error correction protocol for loss on CV states based on CV teleportation and entanglement distillation via noiseless linear amplification. The error correction protocol is concatenated to preserve the same effective transmission coefficient for the quantum channel over increasing distance. The probability of successful operation of the repeater scales polynomially with distance. However, the protocol is limited by a trade-off between fidelity and probability of success.

\end{abstract}

\maketitle

\section{Introduction}
Quantum communication enables various cryptographic protocols that outperform their classical counterparts including Quantum Key Distribution (QKD), with its promise of absolutely secure transmission of information \cite{gisin2002quantum}. The use of quantum optical systems as information carriers is currently the only practical approach to quantum communication  \cite{bachor2004guide}. Never-the-less, one of the biggest challenges facing the realisation of long distance quantum communication is optical loss due to fibre or free-space attenuation. One proposed method to enable long distance transmission of quantum states is the quantum repeater \cite{briegel1998quantum}. In this model, a lossy quantum channel is segmented into smaller, more manageable attenuation lengths along which entanglement is distributed and then purified. Entanglement swapping operations are then performed resulting in entanglement being held between both ends of the quantum channel. 

There have been a number of proposals for quantum repeaters that work on discrete variable quantum systems such as the polarization of single photons \cite{sangouard2011quantum}, and some elements of these have been implemented experimentally. However, quantum communication protocols  can also be implemented using quantum continuous variables \cite{weedbrook2012gaussian}. To date, a complete quantum repeater protocol for continuous variables has not been described, although evidence that CV quantum repeaters can increase transmission distances has been presented \cite{campbell2013continuous} and hybrid protocols combining continuous and discrete states have been proposed \cite{van2008quantum}. It is known that regenerative stations containing only Gaussian elements cannot act as CV quantum repeaters \cite{namiki2014gaussian}.  

In this paper, we outline an architecture for a quantum repeater that may be used with continuous variable quantum optical systems. Our model relies on concatenated error correction protocols consisting of continuous variable teleportation \cite{braunstein1998teleportation} and entanglement distillation via noiseless linear amplification \cite{ralph2009nondeterministic}. The paper is arranged in the following way. In the next section we review the continuous variable error correction protocol that lies at the heart of our repeater. In section III we will describe how the error correction can be concatenated in such a way that the same effective transmission coefficient is maintained even though the physical channel is growing in length. We show that the overhead for this concatenation is polynomial in the length of the channel. We also derive a lower bound for the fidelity of the channel as a function of the channel length. In section IV we evaluate the performance of the continuous variable quantum repeater assuming noiseless linear amplification is implemented via the generalized quantum scissor approach. In section V we consider alternative approaches to implementing the noiseless linear amplification before concluding in the final section.

\section{The Error Correction Protocol}
We use the error correction protocol for continuous-variable states described in Ref~\cite{ralph2011quantum}. This technique for quantum error correction is effective against Gaussian noise induced by loss and proceeds by distilling entanglement and using this entanglement for teleportation. 
\begin{figure}
\centering
\subfigure[]{\label{fig:loss-channel}\includegraphics[width=\linewidth]{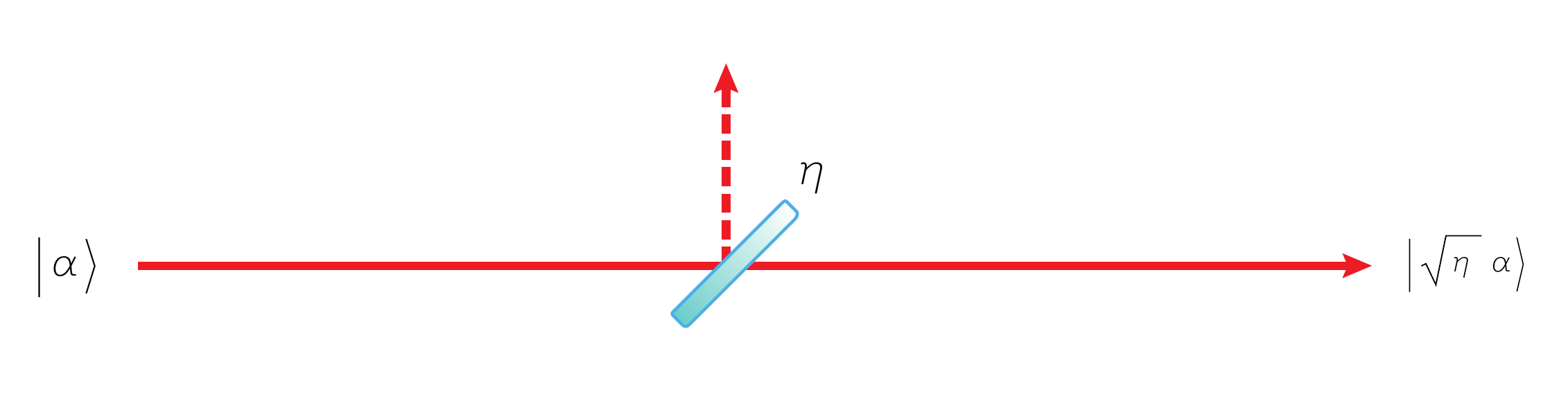}}
\subfigure[]{\label{fig:error-correction1}\includegraphics[width=\linewidth]{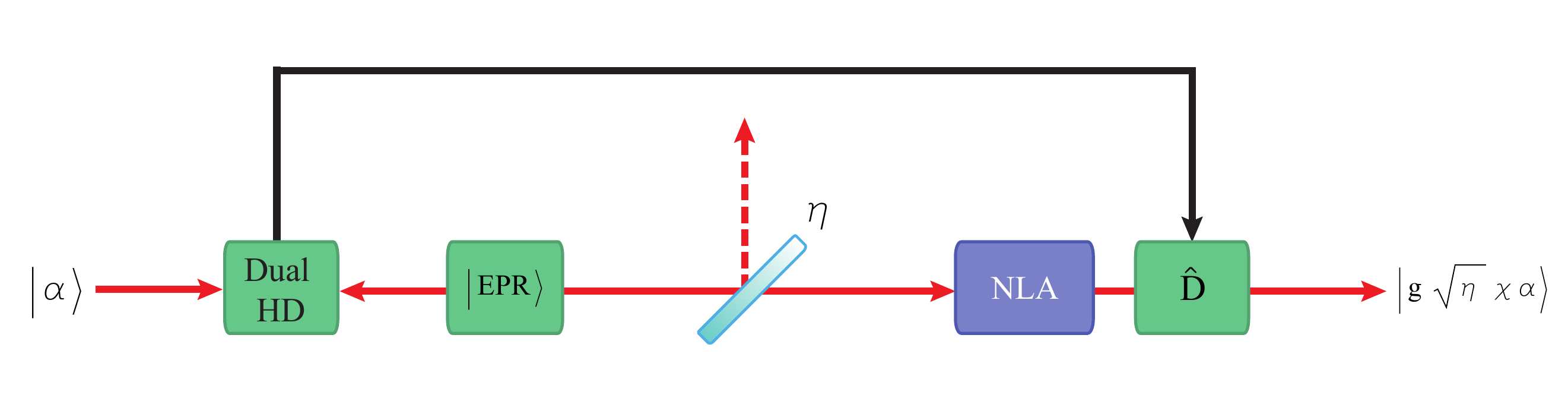}}
\caption{\subref{fig:loss-channel} Lossy channel  \subref{fig:error-correction1} Protocol for quantum error correction from Ref~\cite{ralph2011quantum}. Here EPR entanglement is distributed through a lossy channel of transmission \(\eta\). Noiseless linear amplification is performed to distill the entanglement which is then used for teleportation.}
\label{fig:error-correction}
\end{figure}

The aim of the protocol is to improve the effective transmission of a quantum state through a lossy channel (Fig.1(a)). The protocol is pictured in Fig.\ref{fig:error-correction1} where an Einstein, Podolsky, Rosen (EPR or two mode squeezed) state is distributed through the lossy channel. Distillation is achieved via noiseless linear amplification (NLA) \cite{ralph2009nondeterministic} which is non-deterministic but heralded. When successful, the effect of the NLA on the entanglement is to produce an EPR state of higher purity (for a given entanglement strength) than achievable via direct transmission through the channel. After successful operation of the NLA, the distilled entanglement is used for teleportation: the input signal and the arm of the entangled state that did not pass through the loss are mixed on a 50:50 beamsplitter and conjugate quadratures are detected on each output mode via homodyne detection (also known as dual homodyne detection); the results of the measurement are sent via a classical channel to the receiver; and amplitude and phase modulation proportional to the measurement result are performed to displace the arm of the entanglement that passed through the loss and the NLA, producing the output mode.

For an input coherent state \(\ket{\alpha}\), the action of the lossy channel causes the  transformation:
\begin{equation}
\ket{\alpha}\to\ket{\sqrt{\eta}\alpha}
\label{eq:transform loss}
\end{equation}
where \(\eta\) is the transmission of the channel. In contrast, if the input coherent state is instead teleported using the distilled EPR state and applying gain tuning \cite{polkinghorne1999continuous} we obtain the transformation:
\begin{equation}
\ket{\alpha}\to\ket{g\sqrt{\eta}\chi\alpha}
\label{eq:transform}
\end{equation}
where \(g\) is the gain of the NLA, and \(\chi\) is the strength of the entanglement. By controlling the gain of the NLA, the effective transmission of the channel can be controlled. In particular, we will be interested in the case where \(g\) is chosen to be \(\frac{1}{\eta^{1/4}\chi}\) and the output \eqref{eq:transform} of the protocol is \(\ket{\eta^{1/4}\alpha}\). That is, the channel of transmission \(\eta\) has been error corrected to an effective transmission of \(\eta_{eff}=\sqrt{\eta}\). 

However, it is important to note that the transformation \eqref{eq:transform} is only achieved when the NLA operates in an unphysical asymptotic limit. When implemented with linear optics, the NLA can be constructed from an array of \(N\) modified quantum scissors devices \cite{pegg1998optical}. The input state is split evenly among the \(N\) quantum scissors devices and the state is truncated in the photon number basis to order \(N\). This inevitably limits the fidelity between the input and output states. Additionally, the operation of the NLA is probabilistic and the success probability decreases exponentially with the number of quantum scissors.
 
Never-the-less, as is shown in Ref~\cite{ralph2011quantum}, this protocol can still be effective at correcting errors induced by loss on field states in the high loss regime.

\section{Concatenation of the Error Correction Protocol}
We now present a way to concatenate these error correction protocols in such a way that the effective transmission of the quantum channel is constant with distance and with a probability of success that scales polynomially with distance. In this section we will place bounds on the fidelity of the output state from the repeater based on an assumed fidelity for each of the basic error correction modules. In the subsequent section we will calculate the value for this fidelity under various conditions, and hence estimate the performance of the entire repeater protocol.

\begin{figure}
\centering
\subfigure[]{\label{fig:Concatenate1}\includegraphics[width=0.7\linewidth]{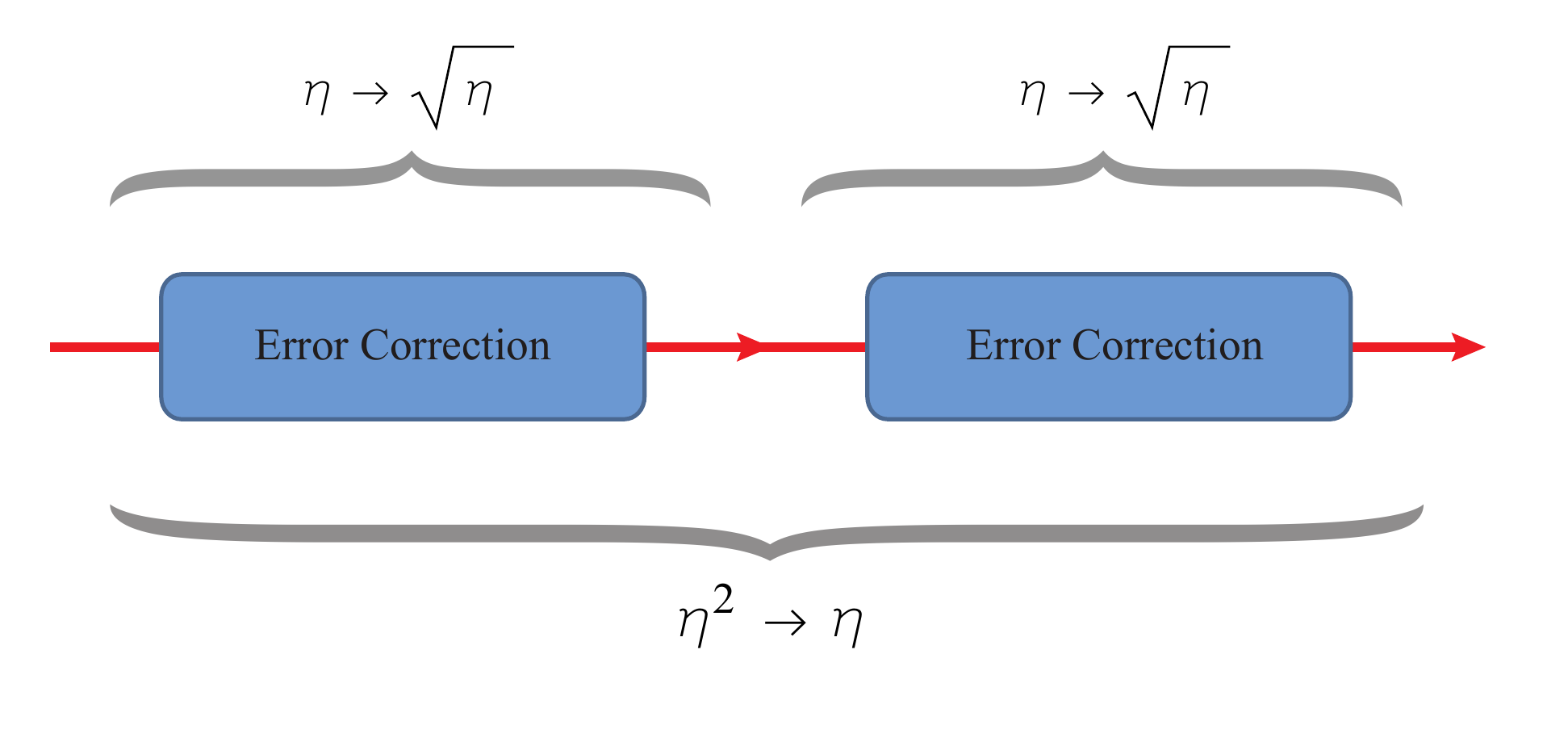}}
\subfigure[]{\label{fig:Concatenate2}\includegraphics[width=0.8\linewidth]{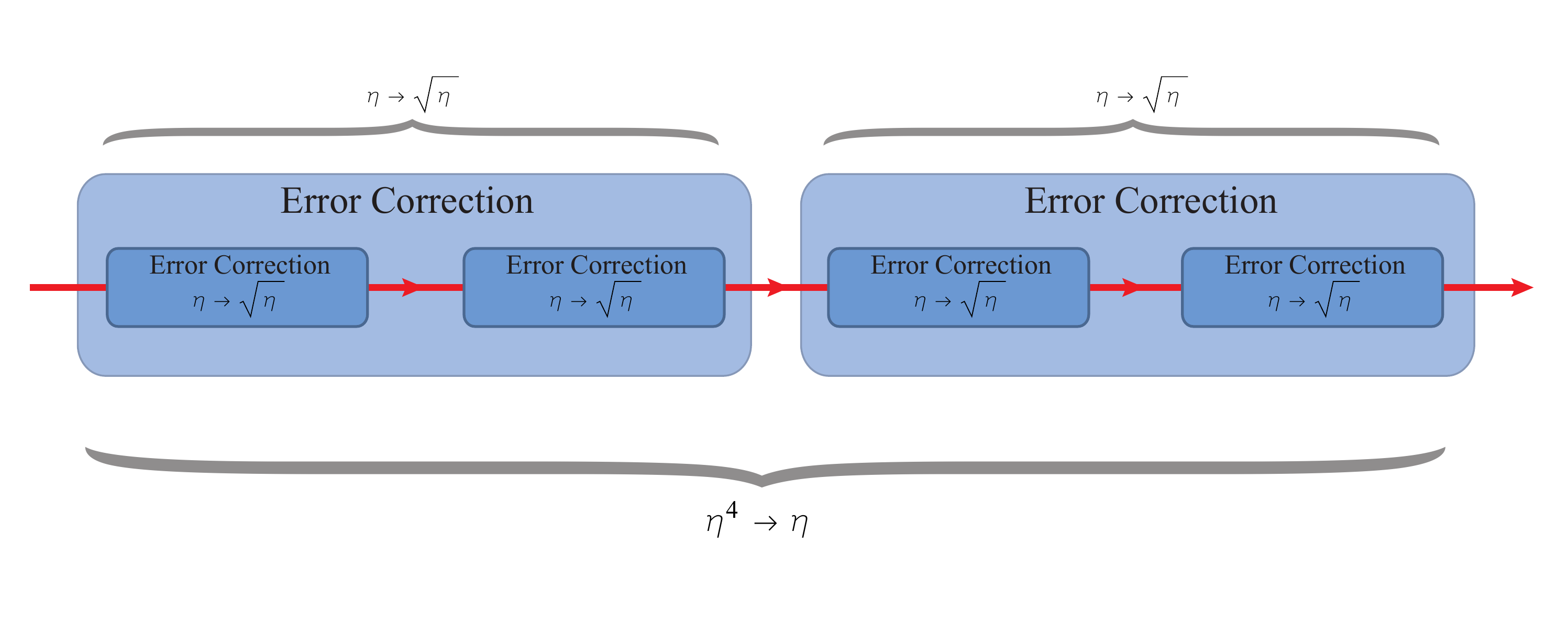}}
\subfigure[]{\label{fig:Concatenate3}\includegraphics[width=\linewidth]{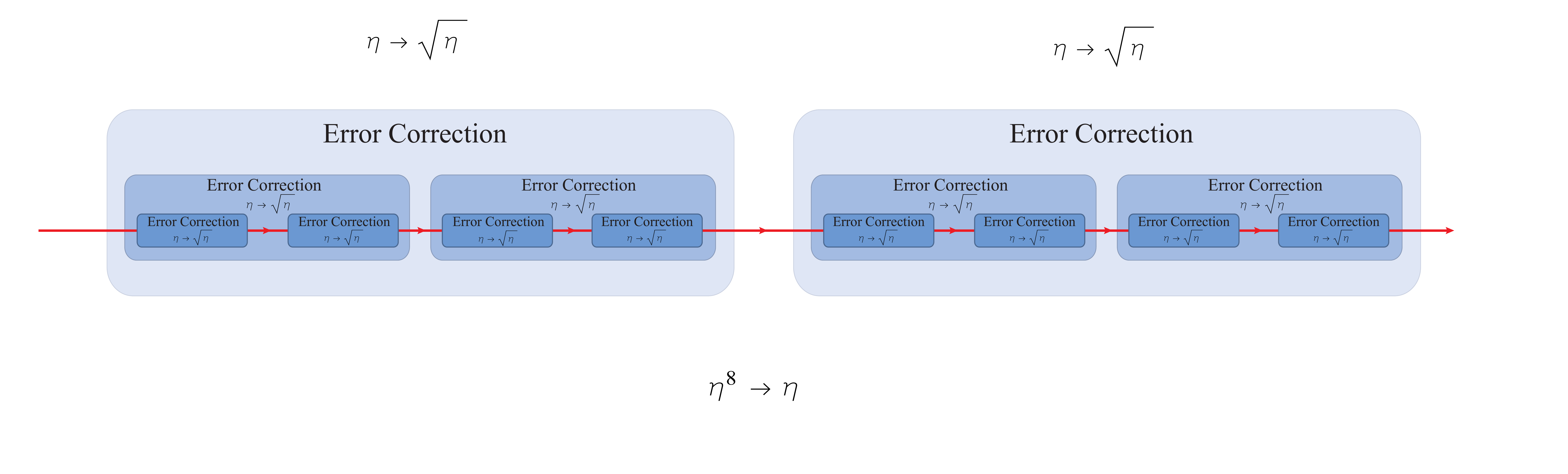}}
\caption{Structure of the quantum repeater for continuous variable states. \subref{fig:Concatenate1} Two links of the repeater. Each box labelled ``error correction" corresponds to the error correction protocol depicted in Fig.~\ref{fig:error-correction1}. \subref{fig:Concatenate2} Four links of the repeater. Nesting two error correction boxes inside a larger error correction box represents the replacement of the physical lossy channel within Fig.~\ref{fig:error-correction1} with the error corrected channel depicted in Fig.~\ref{fig:Concatenate1}. \subref{fig:Concatenate3} Eight links of the repeater. The nesting of error correction modules is concatenated again.}
\label{fig:Concatenation}
\end{figure}

The repeater is depicted in Figure~\ref{fig:Concatenation} where each error correction segment represents the protocol in Fig.~\ref{fig:error-correction1}. Each individual error correction segment takes the initial transmission of the channel \(\eta\) to an effective transmission \(\sqrt{\eta}\). When two error correction protocols are run in series as in Fig.~\ref{fig:Concatenate1}, a quantum channel of overall transmission \(\eta^2\) now has effective transmission \(\eta\). 

To preserve this transmission \(\eta\) over double the distance, another two links of the repeater are necessary as in Fig.~\ref{fig:Concatenate2}. Furthermore, the four base level error correction protocols are nested within two higher level error correction protocols allowing the transmission \(\eta^4\) to appear as effective transmission \(\eta\). If the distance is to be doubled again, another level of concatenation is required as shown in Fig.~\ref{fig:Concatenate3}. Concatenation proceeds in this way for increasing distance where a channel of transmission \(\eta^{2^k}\) requires \(k-1\) levels of concatenation. 

When run in series, two error correction protocols may operate their NLAs independently and simultaneously. Throughout this paper we implicitly assume that high quality quantum memories are available that can store quantum states without loss of fidelity till the synchronising signals arrive from the various NLAs. Therefore, if \(P\) is the success probability for one iteration of the error correction protocol, then the entire protocol in Fig.~\ref{fig:Concatenate1} also operates with success probability \(P\). However, at the first level of concatenation (Fig.~\ref{fig:Concatenate2}) the four individual error correction procedures need to herald successful operation before error correction at the next level of concatenation can proceed. The probability of success for the repeater protocol in Fig.~\ref{fig:Concatenate2} is therefore \(P^2\). Similarly, the success probability for the protocol in Fig.~\ref{fig:Concatenate3} is \(P^3\). Whilst the probability of success is dropping exponentially with the number of concatenations, the distance doubles with each concatenation. Thus, in general we have:
\begin{equation}
P_M = P^{\log_2 M} =  M^{\log_2 P}
\end{equation}
where \(M\) is the number of links of the quantum repeater, and thus a polynomial scaling of success probability with distance.  

To quantify how the quality of the transmitted state decays with distance we use the fidelity, $F$, between input and output states. As stated earlier, the transformation \(\ket{\alpha}\to\ket{g\sqrt{\eta}\chi\alpha}\) is only achieved for fidelity \(F<1\) due to state truncation from the NLA. Formally, the output after one segment of error correction is:
\begin{equation}
\hat{\rho} = F \ket{g\sqrt{\eta}\chi\alpha}\bra{g\sqrt{\eta}\chi\alpha}+(1-F)\hat{\rho}_{\tilde{T}}
\end{equation}
where \(\hat{\rho}_{\tilde{T}}\) is orthogonal to the target state \(\ket{g\sqrt{\eta}\chi\alpha}\). Another iteration of the error correction protocol performs the transformation:
\begin{equation}
\hat{\rho}^\prime = F(F \ket{g^2\eta\chi^2\alpha^2}\bra{g^2\eta\chi^2\alpha^2}+(1-F)\hat{\rho}_{\tilde{T}^2}) + (1-F) \hat{\tilde{\rho}}
\end{equation}
where \(\hat{\rho}_{\tilde{T}^2}\) is orthogonal to the new target state \(\ket{g^2\eta\chi^2\alpha^2}\) and \(\hat{\tilde{\rho}}\) is orthogonal to \(\hat{\rho}\). In this way, two error correction protocols in series (Fig.~\ref{fig:Concatenate1}) produce the required target state \(\ket{g^2\eta\chi^2\alpha^2}\) with fidelity of at least \(F^2\). With two error correction protocols nested within another error correction protocol, the fidelity would be at least \(F^3\) and therefore, the entire protocol in Fig.~\ref{fig:Concatenate2} would have fidelity of at least \(F^6\). In this way, we may say that for \(M\) links of the quantum repeater, the fidelity \(F_M\) between input and output states is bounded below by:
\begin{equation}
F_M \geq F^{2(M-1)}
\label{eq:FM}
\end{equation}

\section{Results} 
In the previous section we described the design of the CV quantum repeater and the scaling properties with distance of its probabilty of success and fidelity, in terms of the probability of success and fidelity of a single error correction module. This fundamental fidelity, \(F\), and success probability, \(P\), are dependent on the entanglement strength of the two mode squeezed state, \(\chi\), and the transmission of the channel between nodes, \(\eta\). Additionally, a higher number of quantum scissors devices employed in the NLA would increase the fidelity, but unfortunately decrease the success probability of the protocol. The task therefore becomes optimising \(F\) and \(P\) to produce the best performance for this quantum repeater.
\begin{figure}
\centering
\includegraphics[width=0.8\linewidth]{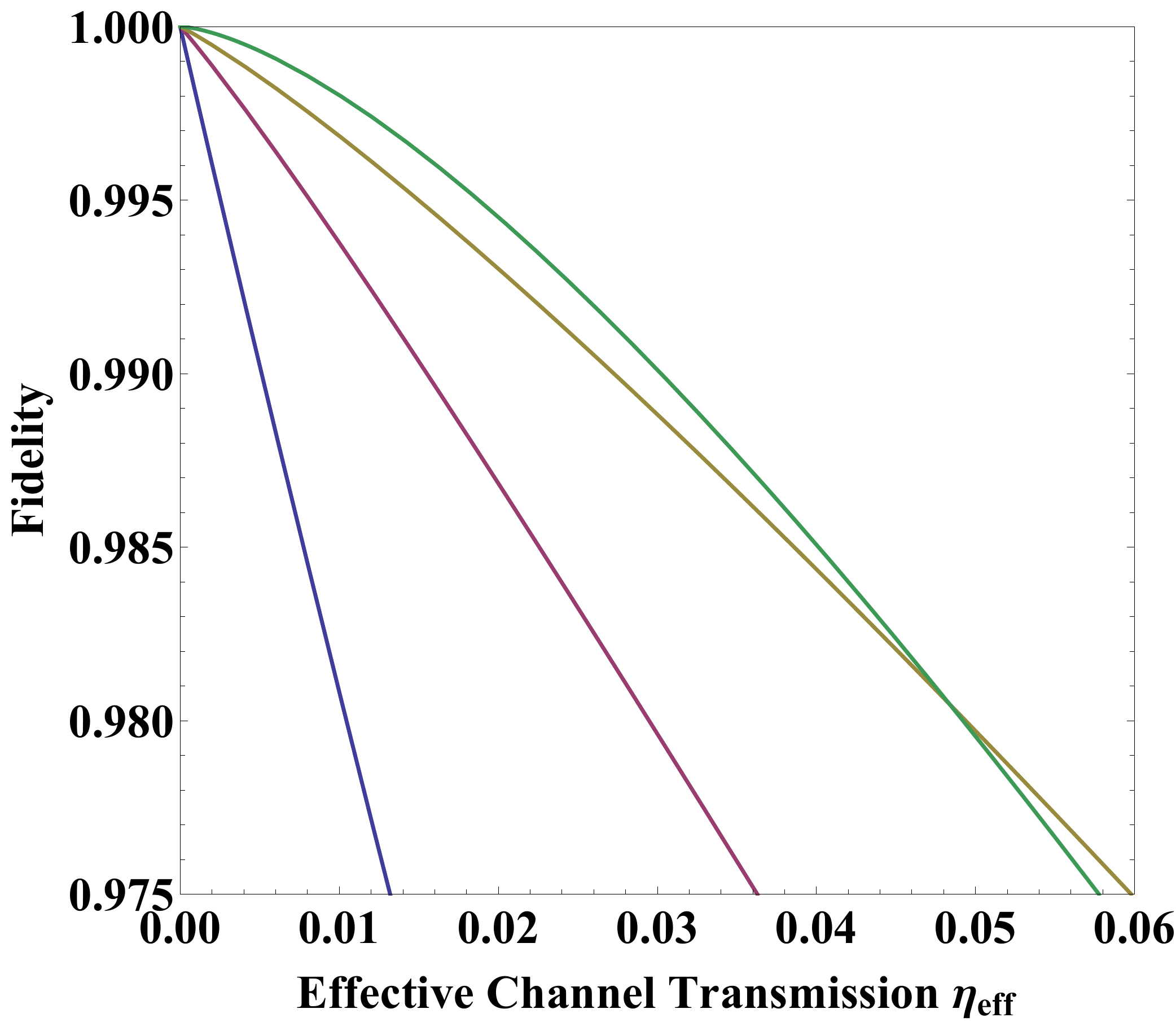}
\caption{Maximum achievable fidelity as a function of effective channel transmission for two links of the quantum repeater (\(\eta^2\to \eta\)). Plotted curves in blue, red and yellow are for the protocol when the NLA is operating with one, two and three quantum scissors respectively. The green line refers to the NLA operation discussed in Section \ref{sec:opt}.}
\label{fig:FidelityM2}
\end{figure}

\begin{figure}
\centering
\includegraphics[width=0.8\linewidth]{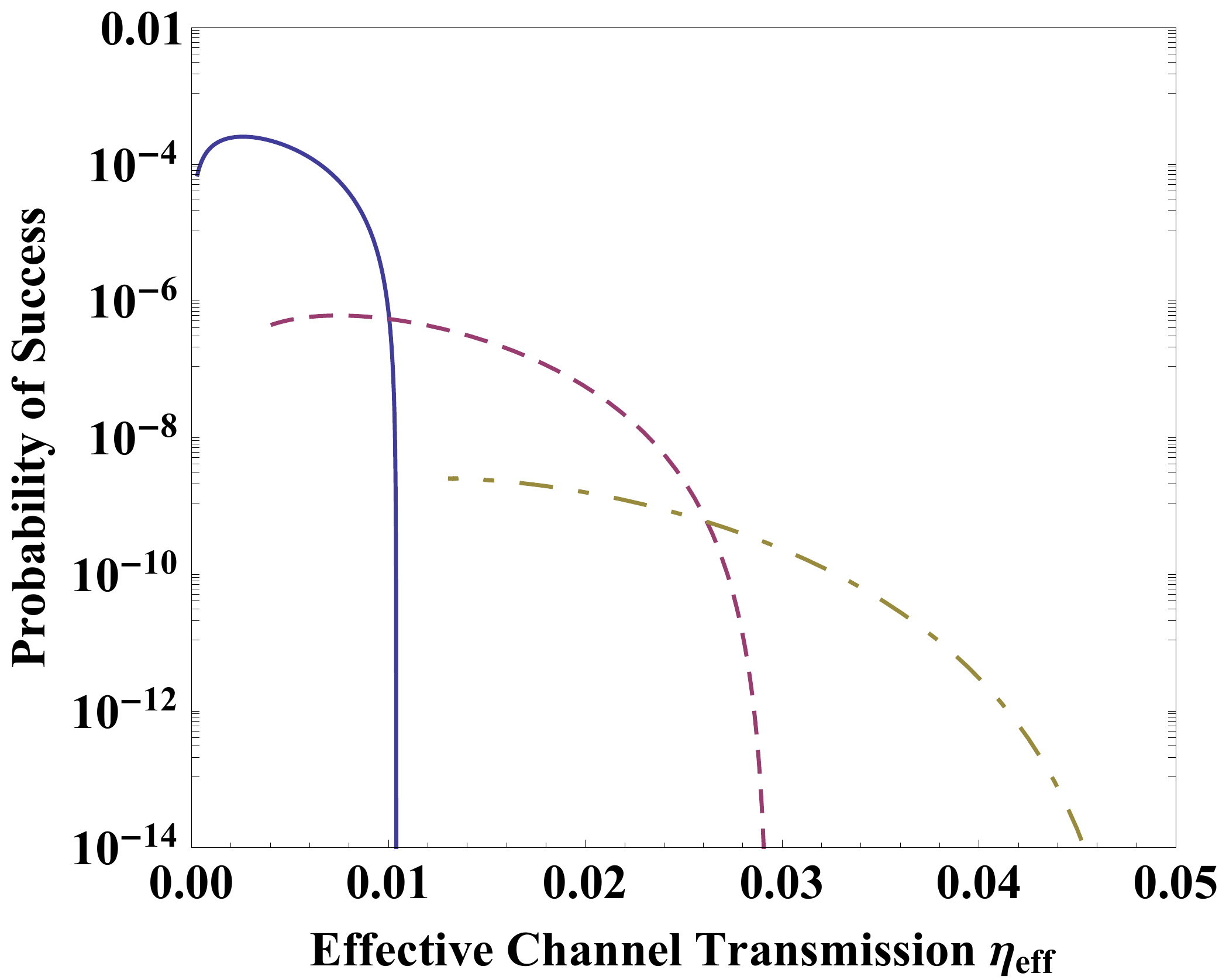}
\caption{Log plot of probability of success of the error corrected channel as a function of effective channel transmission for two links of the quantum repeater (\(\eta^2\to \eta\)). All curves achieve constant fidelity of \(F=0.99\).  The solid line shows the success probability of the protocol operating with a single quantum scissor in the linear optics implementation of the NLA, the dashed line is for two quantum scissors and the dot-dashed line is for three quantum scissors.}
\label{fig:LogPSuccess}
\end{figure}

In the Appendix we detail the calculation of \(F\) and \(P\) assuming that the NLA is implemented using the generalised quantum scissor protocol. We further assume ideal detectors, and single photon and EPR sources. These results are used in the following to examine the performance limits of the CV quantum repeater. 
 
The results contained in Fig.~\ref{fig:FidelityM2} show the maximum achievable fidelity for the two links of the quantum repeater protocol pictured in Fig.~\ref{fig:Concatenate1}. This protocol preserves the effective transmission of a channel \(\eta\) over double the actual distance \(\eta^2\) and the plot shows the fidelities that can be achieved when the error correction protocol uses an NLA that consists of one, two or three quantum scissors. As is evidenced by this plot, using more quantum scissors enables you to achieve higher fidelities that may be impossible with fewer quantum scissors devices. The cost is an exponential decrease in probability of success with increasing numbers of quantum scissors.  

The plot in Fig.~\ref{fig:LogPSuccess} compares the probability of success for two links when the NLA is operating with one, two or three quantum scissors. All plotted curves in Fig.~\ref{fig:LogPSuccess} achieve constant fidelity of \(F = 0.99\). As is shown, with more quantum scissors devices, achieving this high fidelity becomes possible at higher transmissions. However, the success probability decreases significantly in these cases.

\begin{figure}
\centering
\includegraphics[width=0.8\linewidth]{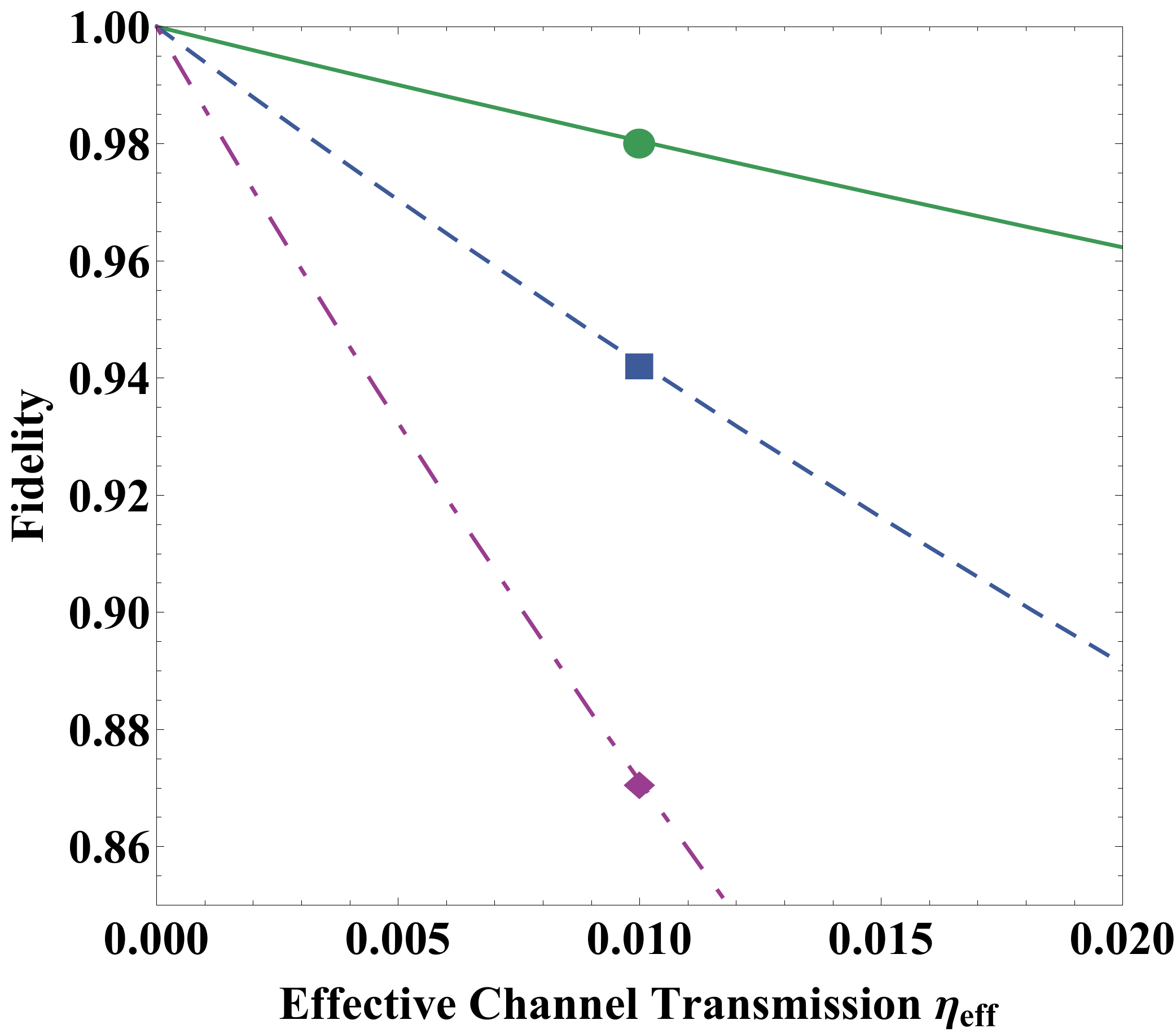}
\caption{Fidelity of the quantum repeater as a function of the effective channel transmission with the error correction protocol using a single quantum scissor and \(\chi = 0.1\). The solid line shows the fidelity for two links of the repeater (\(\eta^2\to\eta\)), the dashed line is for four links of the repeater (\(\eta^4\to\eta\)) and the dot-dashed line is for eight links (\(\eta^8\to\eta\)).}
\label{fig:Fidelity1QS}
\end{figure}
\begin{table}
 \caption{Fidelity and probability of success estimates for varying distances with the quantum repeater. Points correspond to plot in Fig.~\ref{fig:Fidelity1QS}. Also shown in this table are the corresponding fidelities and success probabilities for the protocol when operating with two quantum scissors.}
\begin{center}
    \begin{tabular}{| c | c | c | c | c | c |}
    \hline
    & & \multicolumn{2}{|c|}{One QS} & \multicolumn{2}{|c|}{Two QS} \\ \hline
     & Distance & \(F_M\) & \(P_M\) & \(F_M\) & \(P_M\) \\ \hline
    \textcolor[rgb]{0.24,0.6,0.33692}{\ding{108}} & \( \sim 200 \mathrm{km}\)& 0.98 & 0.001 & 0.99& \(1.1\times 10^{-6}\) \\ \hline
    \textcolor[rgb]{0.24,0.353173,0.6}{\ding{110}} & \(\sim 400\mathrm{km}\)& 0.94 & \(1.2\times10^{-6}\) & 0.98  &\(1.2 \times 10^{-12} \) \\ \hline
    \textcolor[rgb]{0.6,0.24,0.563266}{\ding{117}} & \(\sim 800\mathrm{km}\)& 0.87 & \(1.3\times10^{-9}\) & 0.97 &\(1.3 \times 10^{-18}\) \\
    \hline
    \end{tabular}
\end{center}
\label{tab:Distances}
\end{table}

We now examine the performance of our quantum repeater over varying distances using the fidelity and probability of success. Using reasonable parameters, the plot in Fig.~\ref{fig:Fidelity1QS} shows an example of the fidelity that can be expected from the repeater protocol (where the NLA consists of only a single quantum scissor). The two-mode squeezed state has fixed entanglement strength of \(\chi = 0.1\) and the plot gives fidelity as a function of transmission between repeater nodes. The different lines on this plot give the fidelity for the different number of nodes used in the repeater. 

With the assumption of a loss rate of 0.02dB per kilometre, we note that transmission of \(0.01\) corresponds to loss after approximately 100km of optic fibre. The points highlighted in the plot in Fig.~\ref{fig:Fidelity1QS} correspond to the fidelities with which you can preserve this effective transmission over longer actual distances. Specifically, these distances are 200km, 400km and 800km. Table~\ref{tab:Distances} gives numerical estimates for how the fidelity and probability of success decrease over these distances for the repeater when operating with one or two quantum scissors.

To improve the fidelity you can achieve with this repeater, there are two main options: distribute a weakly entangled EPR state (\(\chi\ll 1\)) and use a high gain of the NLA (\(g\gg 1\)) or employ more quantum scissors in the implementation of the NLA. Both of these options come with the unfortunate cost of a reduction in the probability of success. This signifies the most prominent limitation in using this repeater; that is a high fidelity comes at the expense of the probability of success.

\section{Error Correction with Optimal Amplification \label{sec:opt}}
The trade-off between probability of success and fidelity can be improved by considering more general versions of the NLA than the quantum scissor implementation. For example, suppose we could implement the transformation:
\begin{align}
\ket{0} &\to\ket{0} \nonumber\\
\ket{1} &\to g\ket{1} \nonumber\\
\ket{2} &\to g^2 \ket{2}
\label{eq:Ideal}
\end{align}
This is unlike the transformation of the NLA when implemented with two quantum scissors (which performs \(\ket{0}\to\ket{0}\), \(\ket{1}\to g\ket{1}\) and \(\ket{2}\to\frac{1}{2}g^2 \ket{2}\) up to a normalisation factor). The implementation of the transformation in equation \ref{eq:Ideal} would have an improvement on the fidelity, specifically shown by the green line in Fig.~\ref{fig:FidelityM2}. Here, it can be seen that the fidelity achievable using \eqref{eq:Ideal} is comparable to that of the NLA consisting of three quantum scissors. A physical implementation of eq.~\eqref{eq:Ideal} has been proposed using linear optics \cite{lund2014private}, however the probabilities of a successful transformation are orders of magnitude below its quantum scissors counterpart. The theoretical maximum for probability of success of noiseless amplification has been shown to scale as \(g^{-2N}\) where \(N\) is the order of state truncation \cite{pandey2013quantum}. For the transformation given in \eqref{eq:Ideal}, this would be \(g^{-4}\). Amplification of this type has been theoretically modelled in Ref~\cite{mcmahon2014optimal}. Hence, in principle we can achieve fidelities similar to 3 quantum scissors with probabilities similar to 2 quantum scissors in this way. However, note that the explicit construction in Ref~\cite{mcmahon2014optimal} requires non-linear optical interactions.

\section{Conclusion}
In summary, we have proposed a method to concatenate error correction protocols to produce a quantum repeater that works with CV states. The error correction relies on continuous variable teleportation and entanglement distillation through noiseless linear amplification. While teleportation of CV states is advantageous because of its deterministic operation, it also limits the channel transmission improvement achievable between input and output states. Fidelity is limited by the NLA due to the state truncation. However, the use of CV teleportation also means the protocol will work on any field state and is therefore not limited to a particular optical encoding of quantum information. 

The repeater protocol we present here is limited due to the inevitable trade-off between fidelity and probability of success. As such, there remains significant room for improvement with the protocol used for entanglement distillation. It remains an open question as to how the protocol may be amended to produce higher fidelities while maintaining (or improving) the probability of success. 

\begin{acknowledgments}
We thank Remi Blandino and Austin Lund for useful discussions. This research was funded by the Australian Research Council Centre of Excellence for Quantum Computation and Communication Technology (Project No. CE110001027).
\end{acknowledgments}

\section{Appendix}

In this appendix, we provide details on the calculation of the fidelity \(F\), and the probability of success \(P\) of a single error correction module (pictured in Fig.~\ref{fig:error-correction1}). 
The continuous variable teleportation protocol uses EPR entanglement of the form.
\begin{equation}
\ket{EPR} = \sqrt{1-\chi^2}\sum_{n=0}^\infty \chi^n \ket{n}\ket{n}
\label{eq:EPR_state}
\end{equation} 
In the error correction protocol, one arm of this entanglement is mixed with the input signal \(\ket{\alpha}\) and then conjugate quadratures are detected. The state after detection can be described as:
\begin{equation}
\bra{p_a}\bra{x_b} \hat{U}_{BS} \ket{\alpha}_a \ket{EPR}_b
\label{eq:coherent_epr_projection}
\end{equation}
In \cite{blandino2014channel}, it was shown that this dual homodyne measurement can also be expressed using the equivalence
\begin{equation}
\bra{p_a}\bra{x_b} \hat{U}_{BS} \hat{D}_a(\alpha)\propto \bra{p_a} \bra{x_a}\hat{U}_{BS} \hat{D}_b(-\alpha^*)
\end{equation}
That is, with measurements on the \(\hat{X}\) and \(\hat{P}\) quadratures after a beam splitter, a coherent state of amplitude \(\alpha\) incoming on mode \(a\) is equivalent to a displacement of \(-\alpha^*\) on mode \(b\) and the vacuum \(\ket{0}\) entering mode \(a\).
We now use the result from \cite{blandino2014channel} that states \(\bra{p_a}\bra{x_b} \hat{U}_{BS} \ket{0}_a =\frac{1}{\sqrt{\pi}} \bra{\beta} \), or dual homodyne detection corresponds to a projection onto a coherent state \(\bra{\beta}\) where \(\beta = \frac{x+ip}{\sqrt{2}}\). The state can now be written:
\begin{equation}
\frac{1}{\sqrt{\pi}} \bra{\beta} \hat{D}_b(-\alpha^*)\ket{EPR}_{bc}
\end{equation}
Expanding the EPR state in the number basis gives:
\begin{align}
\sqrt{\frac{1-\chi^2}{\pi}}&\bra{\beta}  \hat{D}_b(-\alpha^*) \sum_n \chi^n \ket{n}_b \ket{n}_c \\
=& \sqrt{\frac{1-\chi^2}{\pi}}\bra{\beta}  \hat{D}^\dagger_b(\alpha^*) \sum_n \chi^n \ket{n}_b \ket{n}_c \\
=& \sqrt{\frac{1-\chi^2}{\pi}}\bra{\beta+\alpha^*}  \sum_n \chi^n \ket{n}_b \ket{n}_c \\
=&\sqrt{\frac{1-\chi^2}{\pi}} \sum_n e^{-\frac{1}{2}|\beta^*+\alpha|^2}\frac{(\chi(\beta^*+\alpha))^n}{\sqrt{n!}} \ket{n} \\
 =& \sqrt{\frac{1-\chi^2}{\pi}} e^{\frac{1}{2}|\beta^*+\alpha|^2(\chi^2-1)}\ket{\chi(\beta^*+\alpha)}
\end{align}
This pure coherent state passess through a lossy channel of transmission \(\eta\), and is transformed as:
\begin{equation}
\hat{\rho}= \frac{1-\chi^2}{\pi}e^{|\beta^*+\alpha|^2(\chi^2-1)}\ket{\sqrt{\eta}\chi(\beta^*+\alpha)}\bra{\sqrt{\eta}\chi(\beta^*+\alpha)}
\label{eq:eta_chi_beta_alpha}
\end{equation}
We then use the Noiseless Linear Amplifier to purify the entanglement. In the quantum scissors implementation of the NLA, with \(N\) quantum scissors, an input number state \(\ket{n}\) is transformed as \cite{ralph2009nondeterministic}:
\begin{equation}
\hat{T}_N \ket{n} = \left(\frac{1}{1+g^2}\right)^{\frac{N}{2}} \frac{N!}{(N-n)!N^n}g^n\ket{n}
\label{eq:T_N}
\end{equation}
For a single quantum scissor, this transformation is:
\begin{equation}
\hat{T}_1(\alpha\ket{0} +\beta\ket{1}) = \sqrt{\frac{1}{g^2+1}}(\alpha\ket{0} + g\beta\ket{1})
\label{eq:single_qs}
\end{equation}
with all higher order terms truncated. Therefore, the state after action of the NLA becomes:
\begin{equation}
\sqrt{\frac{1-\chi^2}{1+g^2}} \frac{1}{\sqrt{\pi}} e^{\frac{1}{2}|\beta^*+\alpha|^2(\chi^2-1-\eta\chi^2)}\left(\ket{0}+g\sqrt{\eta}\chi(\beta^*+\alpha) \ket{1}\right)
\label{eq:qs_state}
\end{equation}
The last remaining step in this error correction protocol is a displacement depending on the result from the dual homodyne measurements, \(\beta\) and a scaling by \(g\sqrt{\eta}\chi\) to account for the entanglement, lossy channel and gain of NLA. The displacement operator \(\hat{D}(-g\sqrt{\eta}\chi\beta^*)\) is applied to \eqref{eq:qs_state} and finally we obtain the un-normalised output state of the error correction protocol:
\begin{align}
\ket{\psi_{1}} = \sqrt{\frac{1-\chi^2}{1+g^2}} \frac{1}{\sqrt{\pi}} e^{\frac{1}{2} |\beta^*+\alpha|^2(\chi^2-1-\eta\chi^2)}\hat{D}(-g\sqrt{\eta}\chi\beta^*)\nonumber\\
\left(\ket{0}+g\sqrt{\eta}\chi(\beta^*+\alpha)\ket{1}\right)
\label{eq:alpha_out}
\end{align}
We can now compute the fidelity, or overlap of this output state with the required target state \(\ket{g\sqrt{\eta}\chi\alpha}\). Fidelity is defined as:
\begin{equation}
F( \ket{\varphi},\ket{\psi}) = |\braket{\varphi|\psi}|^2
\end{equation}
Note that the state \(\ket{\psi_1}\) is dependent on the measurement outcome \(\beta\) from the homodyne detection. In computing the fidelity between this state and the target state \(\ket{g\sqrt{\eta}\chi\alpha}\), we need to average over all possible \(\beta\):
\begin{equation}
F(\ket{g\sqrt{\eta}\chi\alpha}, \ket{\psi_{1}}) = \frac{1}{\int \braket{\psi_{1}|\psi_{1}}\mrm{d}^2\beta} \int|\braket{g\sqrt{\eta}\chi\alpha|\psi_{1}}|^2\mrm{d}^2\beta
\end{equation}
where integration over the complex amplitude \(\mrm{d}^2\beta\) denotes integration over the real and imaginary components of \(\beta\). The factor of \(\frac{1}{\int \braket{\psi_{1}|\psi_{1}} \mrm{d}^2 \beta}\) is needed for normalisation, and it is also important to note that the norm of the un-normalised state gives the probability of success:
\begin{equation}
P_{suc} = \int \braket{\psi_{1}|\psi_{1}}\mrm{d}^2\beta
\end{equation}
\begin{widetext}
We now compute the probability of success. 
\begin{align}
\braket{\psi_{1}|\psi_{1}} = \frac{1-\chi^2}{1+g^2}\frac{1}{\pi} & e^{|\beta^*+\alpha|^2(\chi^2-1-\eta\chi^2)}\left(\bra{0}  +g\sqrt{\eta}\chi(\beta+\alpha^*)\bra{1}\right) \hat{D}^\dagger(-g\sqrt{\eta}\chi\beta^*)\nonumber \\& \hat{D}(-g\sqrt{\eta}\chi\beta^*)\left(\ket{0}+g\sqrt{\eta}\chi(\beta^*+\alpha)\ket{1}\right) \nonumber \\
= \frac{1-\chi^2}{1+g^2}\frac{1}{\pi} & e^{|\beta^*+\alpha|^2(\chi^2-1-\eta\chi^2)}\left(\bra{0}  +g\sqrt{\eta}\chi(\beta+\alpha^*)\bra{1}\right)\left(\ket{0}+g\sqrt{\eta}\chi(\beta^*+\alpha)\ket{1}\right) \nonumber \\
= \frac{1-\chi^2}{1+g^2}\frac{1}{\pi} & e^{|\beta^*+\alpha|^2(\chi^2-1-\eta\chi^2)}\left( 1+g^2\eta\chi^2|\beta^*+\alpha|^2 \right) \nonumber \\
\end{align}
Integrating over \(\beta\):
\begin{align}
\int\braket{\psi_{1}|\psi_{1}} \mrm{d}^2\beta & =\nonumber \\ \int_{-\infty}^\infty \int_{-\infty}^\infty &\frac{1-\chi^2}{1+g^2}\frac{1}{\pi}  e^{|\beta^*+\alpha|^2(\chi^2-1-\eta\chi^2)}\left( 1+g^2\eta\chi^2|\beta^*+\alpha|^2 \right) \mrm{d}\,\mrm{Re}(\beta)\; \mrm{d}\,\mrm{Im}(\beta) \nonumber \\ P_{suc}& = \frac{1-\chi^2}{1+g^2}\frac{1+(-1+\eta+g^2\eta)\chi^2}{(1+(-1+\eta)\chi^2)^2}
\end{align}
We now compute the fidelity:
\begin{align}
\braket{g\sqrt{\eta}\chi\alpha|\psi_{1}} &= \sqrt{\frac{1-\chi^2}{1+g^2}} \frac{1}{\sqrt{\pi}} e^{\frac{1}{2} |\beta^*+\alpha|^2(\chi^2-1-\eta\chi^2)}\bra{g\sqrt{\eta}\chi\alpha}\hat{D}(-g\sqrt{\eta}\chi\beta^*)\left(\ket{0}+g\sqrt{\eta}\chi(\beta^*+\alpha)\ket{1}\right) \nonumber \\
&=\sqrt{\frac{1-\chi^2}{1+g^2}} \frac{1}{\sqrt{\pi}} e^{\frac{1}{2} |\beta^*+\alpha|^2(\chi^2-1-\eta\chi^2)}\bra{g\sqrt{\eta}\chi(\alpha+\beta^*)} \left(\ket{0}+g\sqrt{\eta}\chi(\beta^*+\alpha)\ket{1}\right)\nonumber \\
& = \sqrt{\frac{1-\chi^2}{1+g^2}} \frac{1}{\sqrt{\pi}} e^{\frac{1}{2} |\beta^*+\alpha|^2(\chi^2-1-\eta\chi^2-g^2\eta\chi^2)}(\braket{0|0}+g^2\eta\chi^2|\beta^*+\alpha|^2\braket{1|1})\nonumber \\
&= \sqrt{\frac{1-\chi^2}{1+g^2}} \frac{1}{\sqrt{\pi}} e^{\frac{1}{2} |\beta^*+\alpha|^2(\chi^2-1-\eta\chi^2-g^2\eta\chi^2)}(1+g^2\eta\chi^2|\beta^*+\alpha|^2)
\end{align}
\begin{equation}
|\braket{g\sqrt{\eta}\chi\alpha|\psi_{1}} |^2=\frac{1-\chi^2}{1+g^2}\frac{1}{\pi}e^{|\beta^*+\alpha|^2(\chi^2-1-\eta\chi^2-g^2\eta\chi^2)}(1+g^2\eta\chi^2|\beta^*+\alpha|^2)^2
\end{equation}
Integrating over \(\beta\):
\begin{multline}
\int |\braket{g\sqrt{\eta}\chi\alpha|\psi_{1}} |^2\mrm{d}^2\beta = \\ \int_{-\infty}^\infty \int_{-\infty}^\infty \frac{1-\chi^2}{1+g^2}\frac{1}{\pi}e^{|\beta^*+\alpha|^2(\chi^2-1-\eta\chi^2-g^2\eta\chi^2)}(1+g^2\eta\chi^2|\beta^*+\alpha|^2)^2 \mrm{d}\,\mrm{Re}(\beta)\; \mrm{d}\,\mrm{Im}(\beta) \nonumber
\end{multline}
\begin{align}
F( & \ket{g\sqrt{\eta}\chi\alpha}, \ket{\psi_{1}}) = \frac{1}{\int \braket{\psi_{1}|\psi_{1}}\mrm{d}^2\beta} \int|\braket{g\sqrt{\eta}\chi\alpha|\psi_{1}}|^2\mrm{d}^2\beta \nonumber \\
&= \frac{(1+(-1+\eta)\chi^2)^2}{1+(-1+\eta+g^2\eta)\chi^2} \frac{\left(2 \chi ^2 \left(2 \eta  g^2+\eta -1\right)+\chi ^4 \left(\eta  \left(5 \eta  g^4+4 (\eta -1) g^2+\eta -2\right)+1\right)+1\right)}{\left(\chi ^2 \left(\eta  g^2+\eta -1\right)+1\right)^3} \nonumber \\
&= \frac{\left(1+(-1+\eta ) \chi ^2\right)^2 \left(1+2 \left(-1+\eta +2 g^2 \eta \right) \chi ^2+\left(1+\eta  \left(-2+4 g^2 (-1+\eta )+\eta +5 g^4 \eta \right)\right) \chi ^4\right)}{\left(1+\left(-1+\eta +g^2 \eta \right) \chi ^2\right)^4}
\end{align}
\end{widetext}
Thus we have calculated the fidelity \(F\) and probability of success \(P\) of a single error correction module for the protocol when operating with a single quantum scissor. This fidelity is independent of coherent amplitude and is therefore valid for a coherent state of any amplitude and also an ensemble of coherent states.

Results for two and three quantum scissors are derived similarly, with the only difference being the transformation \eqref{eq:T_N} is applied with \(N=2\) and \(N=3\) respectively. Results for error correction with optimal amplification use \eqref{eq:Ideal} in place of this transformation.

\bibliographystyle{apsrev}
\bibliography{repeater_references}

\end{document}